 \documentstyle{l-aa}             
\font\sixrm=cmr6
  \newcommand{\Msolar}{\mbox{\,$\rm M_{\odot}$}}        
  \newcommand{\Rsolar}{\mbox{\,$\rm R_{\odot}$}}        
  \newcommand{\teff}{\mbox{\,\em T$_{\rm eff}$}}         
  \newcommand\loge{$ \log{\epsilon}$}
  
  \newcommand\loghe{$ \log{\frac{n_{\rm He}}{n_{\rm H}}}$}
  \newcommand{\logg}{\mbox{\,log $g$}}                   
   
  \newcommand{\ion}[2]{\mbox{#1\,{\sixrm #2}}}         
  \newcommand{\vt}{\mbox{\,$v_{\rm t}$}}                 
  \newcommand{\etal}{\mbox{et~al.}}                      
  \def\simge{\mathrel{\raise1.16pt\hbox{$>$}\kern-7.0pt
    \lower3.06pt\hbox{{$\scriptstyle \sim$}}}}           
  \def\simle{\mathrel{\raise1.16pt\hbox{$<$}\kern-7.0pt
    \lower3.06pt\hbox{{$\scriptstyle \sim$}}}}           
\begin{document}

\thesaurus{08.01.1, 08.01.3, 08.08.2, 08.09.2, 08.15.1, 08.18.1}

\title{Spectral analysis of the multi mode pulsating 
subluminous B star PG\,1605+072
\thanks{Based on observations 
    obtained 
    at the W.M. Keck Observatory, which is operated by the Californian 
    Association for Research in Astronomy for the California Institute of 
    Technology and the University of California}
\thanks{based on observations 
    collected at the ESO Southern Observatory}
}

 \author{U. Heber \inst{1}, I.N. Reid \inst{2}, K. Werner\inst{3}
}

   \offprints{U. Heber}

   \institute{Dr. Remeis-Sternwarte, Astronomisches Institut der 
              Universit\"at Erlangen-N\"urnberg,
              Sternwartstr. 7, D\,96049 Bamberg, Germany 
              (E-mail heber@sternwarte.uni-erlangen.de)
              \and
              Palomar Observatory 105-24, California Institute of Technology, 
              Pasadena, CA 91125, USA 
              \and
              Institut f\"ur Astronomie und Astrophysik, Universit\"at T\"ubingen, 
              D\,72076 T\"ubingen
}

    \date{Received ; accepted }

    \maketitle
   \markboth
   {U. Heber \etal: Spectral analysis of the pulsating 
sdB star PG\,1605+072 }
   {U. Heber \etal: Spectral analysis of the pulsating 
sdB star PG\,1605+072 
}

%
%
\begin{abstract}
PG\,1605+072 
has unique pulsational properties amongst
the members of the new class of pulsating sdB (EC\,14026) stars.
It has the longest periods and the richest, 
most puzzling frequency spectrum (55 periods).  
We present a quantitative analysis
of a Keck HIRES spectrum using 
NLTE and LTE model atmospheres.  
Atmospheric parameters (\teff, \logg, \loghe), metal abundances, 
and the rotational velocity
are determined. He, C, N and O are subsolar as well as the 
intermediate mass elements Mg and Si, whereas Ne and Fe are solar. 
This abundance pattern is caused by diffusion. PG\,1605+072
displays considerable line broadening probably caused by rotation 
(v\,sini = 39\,km/s, P$<$8.7h) and, therefore, is 
predicted to evolve into an unusually fast rotating white dwarf.
Unequal rotational splitting may explain its puzzling 
pulsation pattern. The solar Fe abundance and the rapid rotation nicely confirms
recent predictions of diffusion/pulsation models.


\end{abstract}
%
\section{Introduction}

Hot subluminous B 
stars (sdB) form a homogeneous group dominating the population of faint blue 
stars (B$<$16$^{\rm m}$).
Following ideas outlined by Heber (1986) the sdB 
stars can be identified with 
models
for extreme HB (EHB) stars, 
which differ markedly from those for normal HB stars. 
An EHB star bears great resemblance to a helium main-sequence star 
of half a solar mass and its further evolution should proceed similarly (i.e. 
directly to the white dwarf graveyard,
Dorman et al. 1993).

Recently, 
several sdB stars have been found to be pulsating (termed EC14026 stars 
after the prototype, see O'Donoghue et al. 1999 
for a review), 
defining a new instability strip in the HR-diagram. 
The study of these pulsators 
offers the possibility of applying  
the tools of asteroseismology to investigate the structure of sdB stars.
The existence of pulsating sdB stars was predicted by 
Charpinet et al. (1996), who uncovered an efficient driving mechanism due 
to an opacity bump associated with iron ionization in EHB models. However,
in order to drive the pulsations, iron needed to be enhanced in the appropriate 
subphotospheric layers, possibly due to diffusion. Subsequently, 
Charpinet et al. (1997) confirmed this assumption by detailed
diffusion calculations. Even more 
encouraging was the agreement of the observed and predicted instability 
strip. 

Thirteen pulsating sdB stars are well-studied photometrically (O'Donoghue 
et al. 1999).
A precise knowledge of effective temperature, gravity, element 
abundances and rotation is
 a prerequisite for the asteroseismological investigation.

PG\,1605+072 was selected for a detailed quantitative spectral analysis because 
it displays the richest 
frequency spectrum amongst the EC\,14026 stars ($>$50 periods have been 
identified, Kilkenny al. 1999). Recently, Kawaler (1999) predicted from his 
modelling of the pulsations that PG\,1605+072 should be rotating.

\section{Observation and data reduction}

A high resolution optical spectrum
of PG\,1605+072 was obtained with the 
HIRES echelle spectrograph (Vogt et al. 1994) on the Keck 
I telescope on July 20, 1998 using the blue cross 
disperser to cover the full wavelength region between 3700\AA\ and 5200\AA\
at a resolution 0.09\AA.

The exposure time (600s) is longer 
than the pulsational periods (206--573s).
The standard data reduction as described by Zuckerman \& Reid (1998)
resulted in spectral orders that have a somewhat wavy 
continuum. In order to remove the waviness we used the spectrum of H1504+65 
(a very hot pre-white dwarf devoid of hydrogen and helium, Werner 1991) 
which was observed in the same night. Its spectrum has only few weak lines 
of highly ionized metals in the blue (3600--4480\AA) where the strong Balmer 
lines are found in the sdB stars. Therefore we normalized 
individual spectral orders 1 to 20 (3600--4480\AA) of the sdB stars by
dividing through the smoothed spectrum of H1504+65. The remaining 
orders were normalized by fitting the continuum with spline functions
(interpolated for orders 26 and 27 which contain H$\beta$). 
Judged from the match of line profiles in 
the overlapping parts of neighboring orders this procedure worked 
extremely well. Atmospheric parameters determined from individual
Balmer lines are found to be consistent with each other except for H$\beta$.
Therefore, we excluded H$\beta$ from the fit procedure. 
Moreover, the resulting \teff\ and \logg\ are also in excellent agreement with 
those from the fit of a low resolution spectrum
obtained at the ESO NTT (provided by S. 
Moehler).
Weak
lines of C, N, O, Ne, Mg, Si, and Fe  
can also be identified. 

\section{Atmospheric parameters}

The simultaneous fitting of Balmer and He line profiles by a grid of 
synthetic spectra (see Saffer et al. 1994) has become the standard 
technique to determine the atmospheric parameters of sdB stars.
The Balmer lines (H$\gamma$ to H\,12), \ion{He}{I} 
(4471\AA, 4026\AA, 4922\AA, 
4713\AA, 5016\AA, 5048\AA) and \ion{He}{II} 4686\AA\ lines are fitted to 
derive all three parameters simultaneously. 

The analysis is based on
grids of metal line blanketed LTE model atmospheres  
for solar metalicity and Kurucz'
ATLAS6 Opacity Distribution Functions (see Heber et al. 1999). Synthetic 
spectra are calculated with  
Lemke's LINFOR program (see Moehler et al. 1998).

The results are listed in Table 1 and compared to published values obtained 
from low resolution spectra. The formal errors of our fits 
are much smaller than the systematic errors (see below). 
The agreement with the results from low resolution spectra analysed with 
similar models
(Koen et al. 1998) as well as from our own low resolution spectrum for PG\,1605+072 is very 
encouraging. 

\begin{table}
\caption{Atmospheric parameters for PG\,1605+072 from different methods, 
see text}\label{res_av}
\begin{tabular}{|l|lll|}
\hline
method         & \teff\ [K]      & \logg\        & \loghe \\
\hline
LTE:                 &                 &               &     \\
Koen et al. (1998)   & 32\,100$\pm$1000 & 5.25$\pm$0.1 & --\\
NTT, low res.        & 32\,300& 5.31& -2.46\\
HIRES, H+He          & 31\,900 & 5.29 & -2.54\\
HIRES, He            & 33\,200 & 5.34 & -2.42\\
NLTE:                &                 &               &     \\
NTT, low res.        & 32\,500 & 5.25 & -2.46\\
HIRES, H+He          & 32\,200 & 5.21 & -2.54\\
HIRES, He            & 32\,400 & 5.21 & -2.53\\
\hline
adopted              & 32\,300$\pm$300 & 5.25$\pm$0.05 & -2.53$\pm$0.1\\
\hline
\end{tabular}
\end{table}

\begin{figure*}
\vspace{8.1cm}
\includegraphics{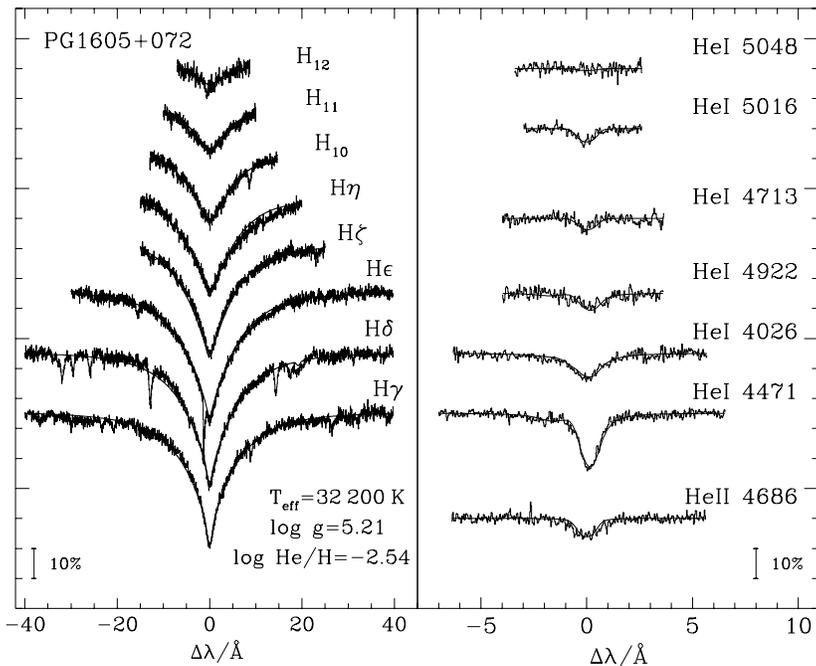}
\caption[]{Balmer and He line profile fits for PG\,1605+072 of the HIRES 
spectrum
from NLTE model atmospheres.}
\end{figure*}

Four species are represented by two 
stages of ionization (\ion{He}{I} and \ion{He}{II}, \ion{C}{II} and 
\ion{C}{III}, \ion{N}{II} and \ion{N}{III}, \ion{Si}{III} and \ion{Si}{IV}). 
Since these line ratios 
are
very temperature sensitive at the temperatures in 
question, we alternatively can 
derive \teff\ and abundances by matching these ionization equilibria.  
Gravity is derived 
subsequently from the Balmer lines by keeping \teff\ and \loghe\ fixed. These
two steps are iterated until consistency is reached. 
\ion{C}{II} is represented by the 4267\AA\ line 
only, which is known to give notoriously too low carbon abundances. Indeed 
the carbon ionization equilibrium can not be matched at any reasonable 
\teff. The ionization equilibria of He, N and Si require \teff\ to be 
higher than from the Saffer procedure, i.e. 33\,200K (He), 33900K (N) and
32\,800K (Si). Table 1 lists the result for the He ionization equilibrium.   

This difference could be caused by NLTE effects. Therefore we repeated the
procedure for \teff\ and \loghe\ using 
a grid of H-He line blanketed, metal free NLTE 
model atmospheres (Napiwotzki 1997), calculated with the ALI 
code of Werner \& Dreizler (1999). NLTE calculations for N and Si are 
beyond the scope of this letter. 

Applying Saffer's procedure with the NLTE model grid (see Fig. 1) 
yields \teff\ almost
identical to that obtained with the LTE grid. 
Evaluating the He ionization equilibrium in NLTE, indeed, results in \teff\
being consistent with that from Saffer's procedure (see Table 1). 
We therefore conclude that 
the higher \teff\ derived above from the ionization equilibrium in LTE 
is due to NLTE 
effects.
 
However, a systematic difference in \logg\ persists, the LTE values
being higher by 0.06 -- 0.08 dex than the NLTE results (see Table 1).
Since its origin is obscure, we finally adopted the 
averaged atmospheric parameters given in Table 1. 
Helium is
deficient by a factor of 30 as
is typical for sdB stars.

\section{Metal abundances}

The metal lines are sufficiently isolated to derive 
abundances from their equivalent widths except for the crowded region 
from 4635\AA\ to 4660\AA\ which we analyse by detailed spectrum synthesis.
Results are listed in Table 2 and plotted in Figure 2.
Although several O lines are available, it was impossible to determine the 
microturbulent velocity(\vt) in the usual way, i.e. by minimizing the
slope in a plot of the O abundances versus equivalent widths.
We adopted \vt = 5$\pm$5km/s which translates 
into small systematic abundance uncertainties of $\pm$0.05dex for most 
ions. The analysis is done in LTE and we therefore used a model (\teff=
33500\,K, \logg = 5.35) that is consistent with the He, N and Si ionization 
equilibria. A temperature uncertainty of $\Delta$\teff=1000\,K translates 
into abundance uncertainties of less than 0.1\,dex. Hence systematic errors 
are smaller for most ions than the statistical errors given in Table 2.

\begin{table}
\caption{Metal abundances for PG\,1605+072 compared to solar composition. n 
is the number of spectral lines per ion.}\label{res_av}
\begin{center}
\begin{tabular}{|l|rll|}
\hline
ion            & n     & \loge     & [M/H] \\
\hline
\ion{C}{III}          & 2   & 7.80$\pm$0.07  & $-$0.83\\
\ion{N}{II}           & 5    & 7.62$\pm$0.16 & $-$0.33  \\
\ion{N}{III}          & 2    & 7.66$\pm$0.24 & $-$0.29\\
\ion{O}{II}           & 25   & 8.00$\pm$0.21 & $-$0.88\\
\ion{Ne}{II}          & 4    & 7.57$\pm$0.19 & $-$0.12\\
\ion{Mg}{II}          & 1    & 7.16  & $-$0.33\\
\ion{Si}{III}         & 3    & 7.11$\pm$0.06 & $-$0.35 \\
\ion{Si}{IV}          & 4    & 7.07$\pm$0.16& $-$0.39\\
\ion{Fe}{III}         & 4    & 7.62$\pm$0.39 & +0.30  \\
\hline
\end{tabular}
\end{center}
\end{table}

Carbon and oxygen are depleted by 0.8--0.9 dex with respect to solar 
composition whereas nitrogen, magnesium and silicon are only slightly 
deficient (factor 2). Surprisingly neon and iron are solar to within error 
limits. This peculiar abundance pattern is probably due to diffusion, i.e. 
the interplay of gravitational settling and radiative levitation.
The iron abundance is of special interest as an diffusive 
enrichment in subphotospheric layers is necessary to drive the pulsations. 
Its surface abundance is 
consistent with the diffusion calculations of Charpinet et al. (1997). 

\begin{figure}
\vspace{4.4cm}
\includegraphics{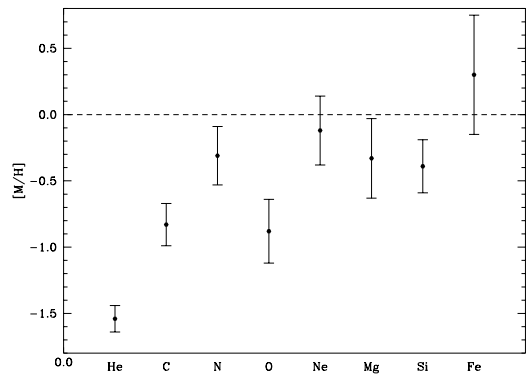}
\caption[]{Abundances of PG\,1605+072 relative to the sun. 
For Mg an 
uncertainty of 0.3\,dex was assumed. 
}
\end{figure}

\section{Projected rotational velocities}

The spectral lines of PG\,1605+072 are considerably broadened,
which we attribute to stellar rotation and derive 
v\,sin\,i = 39km/s, by fitting the strongest metal lines.
In Figure 3 we compare a section of the spectrum of PG\,1605+072 to that of the 
pulsating sdB star Feige\,48 observed with the same 
instrumental setup. However, 
oscillations in these objects should be associated with motions 
which could lead to line broadening similar to the rotation effect.  
For a radial pulsation with a sinusoidal velocity curve a radius change of 
$\Delta$R/R=$\frac{v\,P}{\pi\,R}\ge$3\% 
for PG\,1605+072 would be required. In comparison 
Feige\,48 is very sharp-lined (v\,sin\,i$<$8km/s)
corresponding to $\Delta$R/R$<$0.6\% if sin\,i=1.  
Although the intensity amplitudes of the pulsations in Feige\,48 are 
considerably 
smaller than for PG\,1605+072 we regard rotation as the 
more likely reason for the broadening of the lines of the latter. 
Time resolved spectroscopy 
should allow to disentangle both effects as well as to search for 
temperature changes associated with the pulsations.  

Assuming a mass of 0.5\Msolar\ the radius of R=0.28\Rsolar\ follows from the 
gravity. Since sin\,i cannot be constrained 
the corresponding rotation period of PG\,1605+072 must be smaller than 8.7h.
PG\,1605+072 displays the most complex power spectrum with more than 50
frequencies identifiable (Kilkenny et al. 1999), 
39 being bona fide normal pulsation frequencies.

Usually rotation becomes manifest in the power spectrum by 
the characteristic splitting into 
equidistantly spaced multiplet components as is observed e.g. for 
the pre-white dwarf PG\,1159-035 (rotation period: 1.4\,d, Winget 
et al. 1991). Such multiplet's, however, have not been identified for 
PG\,1605+072. 
Fast rotation introduces higher order terms that result
in unequally spaced multiplet components.
Recently, Kawaler (1999), was able to identify the five main peaks by 
considering mode trapping and rotational splitting.
He predicted that PG\,1605+072 should be rapidly rotating (130\,km/s). The 
measured v\,sin\,i=\,39\,\,km/s, hence, is a nice confirmation of Kawaler's
prediction.

Rotation is interesting also from the point of view of stellar evolution.
PG\,1605+072 is probably already in a post-EHB phase of evolution
(Kilkenny et al. 1999)
and will evolve directly into a white dwarf, i.e. will shrink 
from its present radius of 0.28\Rsolar\ to about 0.01\Rsolar. Hence 
PG\,1605+072 will end 
its life as an unusually fast rotating white dwarf if no loss of angular 
momentum occurs. Isolated white dwarfs, however, are known to be very slow 
rotators (e.g. Heber et al. 1997).

\begin{figure}
\vspace{8.2cm}
\includegraphics{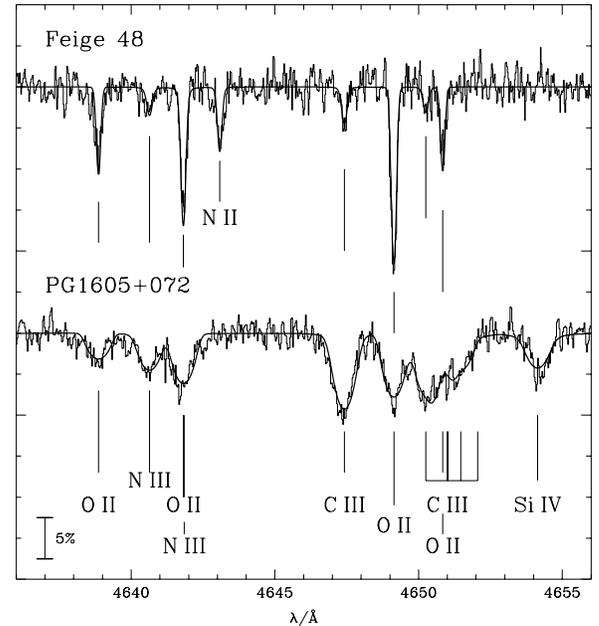}
\caption[]{Fit of a section of the metal line spectrum of
PG\,1605+072, (bottom, v\,sini=39\,km/s) compared to that of Feige\,48 
(top, no rotation), another 
pulsating sdB star.}
\end{figure}

\section{Conclusion}

Atmospheric parameters of PG\,1605+072 and abundances of several elements 
have been determined at the highest accuracy achieved so far for an EC14026
pulsator. Together with the complex pulsation pattern this 
makes PG1605+072 the most interesting object 
of its class. 

The high projected rotational velocity of PG\,1605+072 (v\,sin\,i=39\,km/s) 
may be a key observation to understand its puzzling power spectrum. This 
measurement nicely confirms a recent prediction from pulsational models 
(Kawaler 1999). 

Although the helium abundance is low ($\approx$1/30\,solar) 
as is typical for sdB stars, most metals are only mildly depleted 
(0.3--0.8\,dex) whereas Ne and Fe are solar. The solar iron abundance is in 
perfect agreement with the predictions from the diffusion models 
of Charpinet et al. (1997).
The determination of 
abundances of several other elements provides a serious test of diffusion 
models. 

\acknowledgements  

We thank 
Sabine Moehler for providing us with her low resolution
spectrum of PG\,1605+072 and 
Ralf Napiwotzki and Gilles Fontaine for fruitful discussions. 
A travel grant by the DFG is
gratefully acknowledged.   
\vspace{-0.2cm}
%
%

\end{document}